\documentclass[preprint,showpacs,preprintnumbers,amsmath,amssymb,superscriptaddress]{revtex4}

\usepackage{graphicx,amsfonts}% Include figure files
\usepackage{epsfig}
\usepackage{dcolumn}% Align table columns on decimal point
\usepackage{bm}% bold math
\begin{document}

%\preprint{IFT-P.xx/2009}
%\preprint{ArXiv:yymm.nnnn}
\title{Electron and neutron electric dipole moment in the 3-3-1 model with heavy leptons}

% \altaffiliation[Also at ]{Physics Department, XYZ University.}%

\author{G. De Conto}%
\email{georgedc@ift.unesp.br}
\affiliation{
Instituto  de F\'\i sica Te\'orica--Universidade Estadual Paulista \\
R. Dr. Bento Teobaldo Ferraz 271, Barra Funda\\ S\~ao Paulo - SP, 01140-070,
Brazil
}
\author{V. Pleitez}%
\email{vicente@ift.unesp.br}
\affiliation{
Instituto  de F\'\i sica Te\'orica--Universidade Estadual Paulista \\
R. Dr. Bento Teobaldo Ferraz 271, Barra Funda\\ S\~ao Paulo - SP, 01140-070,
Brazil
}

\date{11/26/14}% It is always \today, today,
\begin{abstract}
We calculate the electric dipole moment for the electron and neutron in the framework of the 3-3-1 model with heavy charged leptons.
We assume that the only source of $CP$ violation arises from a complex trilinear coupling constant and the three complex vacuum expectation values. However, two of the vacua phases are absorbed and the other two are equal up to a minus sign. Hence only one physical phase survives. In order to be compatible with the experimental data this phase
has to be smaller than $10^{-6}$.
\end{abstract}


\maketitle
% % % % % % % % % % % % % % % % % % % % % % % % % % % % % % % % % % % % % % % % % % %
%\documentclass[]{article}
%\usepackage{amsmath}
%\usepackage{ae}
%\usepackage{slashed}
%\usepackage{graphicx}

%opening
%\title{The neutron EDM in the 3-3-1 model with heavy leptons}
%\author{G. De Conto, V. Pleitez}

%\begin{document}

%\maketitle

%\begin{abstract}
%
%\end{abstract}

\section{Introduction}

The measurement of the electric dipole moment (EDM) of elementary particles is a crucial issue to  particle physics. This is because  for a nondegenerate system, as nucleus or  an elementary particle, an EDM is possible only if the symmetries under $T$ and $C\!P$ are violated. On one hand, in the Standard Model (SM) framework the only source of $T$ and $C\!P$ violation is the phase $\delta$ in the CKM mixing matrix.
On the other hand, the SM prediction for the neutron electric dipole moment is $|d_n|_{\textrm{SM}}\approx10^{-32}\,e\cdot$ cm \cite{Shabalin:1978,
Shabalin:1983,Shabalin:1980,Eeg:1984,Czarnecki:1997}, 6 orders of magnitude below the actual experimental limit of $|d_n|_{\textrm{exp}}<2.9\times 10^{-26}\,e\cdot \textrm{cm}$ \cite{Baker:2006}.
Moreover, for the electron EDM the SM prediction of $|d_e|_{\textrm{SM}}<10^{-38}\,e\cdot \textrm{cm}$ \cite{Cummins:1999} and the experimental upper limit of $|d_e|_{\textrm{exp}}<8.7 \times 10^{-29}\,e\cdot \textrm{cm}$ \cite{Baron:2013eja}.
Hence, we see that in the context of the Standard Model the Kobayashi-Maskawa phase is not enough for explaining an EDM with a value near the experimental limit for both electron and neutron.
If the latter case is confirmed in future experiments, it certainly means the discovery of new physics with new $C\!P$ violation sources.

We can rewrite the experimental upper bound of the electron EDM in units of Borh magneton as follows
\begin{equation}
d_e< 5\times 10^{-17}\;\mu_B\sim\frac{2m_e}{M}\, \mu_B,
\label{bohr}
\end{equation}
where $M$ is the particle responsible for the nonvanishing EDM of a given particle. From Eq.~(\ref{bohr})
we obtain that $M>5\times 10^{14}$ GeV. This naive calculation assumes that the electron EDM arises only by the effect of a massive particle. Notwithstanding, in a specific model the masses of the particles responsible for the EDM may be much smaller than this value since it does not take into account the couplings of the responsible particle and negative interference if there are several of such particles. This is the case in electroweak models and, in particular, in the 3-3-1 ones. In the latter models there are many $C\!P$ violating phases like the Kobayashi-Maskawa $\delta$, but these are hard phases in the unitary matrices that relate the symmetry eigenstates and the mass eigenstates that, unlike in the SM, survive in some interactions among quarks, vector bosons and scalars.

Moreover, cosmology also hints that the SM may not be a complete description and that new $CP$ violating phases must exist in models beyond the SM in order to explain the observed matter-antimatter asymmetry of the Universe~\cite{Riotto:1999, Morrisey:2012, Dine:2004}. Therefore, we are led to explore alternatives to the SM; in our case we consider the 3-3-1 model with heavy leptons (331HL for short)~\cite{Pleitez:1992xh}.  However, in this work we will only be concerned with the EDM issue.

The outline of this paper is as follows.
In Sec.~\ref{sec:model} we introduce the representation content of the model: in Sec.~\ref{subsec:scalars} we write the scalar content, in Sec.~\ref{subsec:leptons} the lepton sector, and quarks in Sec.~\ref{subsec:quarks}. In 
Sec.~\ref{sec:edmft} we calculate the EDMs, for the electron in Sec.~\ref{subsec:edmleptons} and  for the neutron in \ref{subsec:neutronedm}. The last section, Sec.~\ref{sec:con}, is devoted to our conclusions. In the Appendixes~\ref{sec:scalars2}~--~\ref{sec:verticesgaugescalar} we write all the interactions used in our calculation.

\section{The 3-3-1 model}
\label{sec:model}

Here we will work in the framework of the 3-3-1 model with heavy leptons proposed in Ref.~\cite{Pleitez:1992xh}.  In this, as in other 3-3-1 models, there are many phases in the mixing matrices. Even if the phases in the CKM mixing matrix are absorbed in the quark fields, they appear in the interactions of the fermions with heavy vector and scalar bosons~\cite{Promberger:2007py}. Here we considered the case when the only $C\!P$ violating phase is that of the soft trilinear interaction  in the scalar potential and the three VEV are also considered complex. However, the phases in the VEVs $v_\eta$ and $v_\rho$, can be rotated away with a $SU(3)$ transformation and the stationary condition imposes a relation between the other two, thus only one physical phase survive. It was shown in Refs.~\cite{Montero:1998yw,Montero:2005yb} that their model has a mechanism for $CP$ violation, but their detailed calculations of the EDMs were not done. This was mainly because at the time we did not know realistic values for the matrices $V^{U,D}_{L,R}$ and $V^l_{L,R}$ their numerical values are given in Sec.~\ref{subsec:leptons}. 
Expressions for the matrices in the quark sector were found in Ref.~\cite{Machado:2013jca} in the context of the nontrivial SM limit of the model found in Ref.~\cite{Dias:2006ns}. See Sec.~\ref{subsec:quarks}.

In this model, as in the minimal 3-3-1, the electric charge operator is given by
\begin{equation}
\frac{Q}{|e|}=T_{3}-\sqrt{3}T_{8}+X,
\end{equation}
where $e$ is the electron charge, $T_{3,8}=\lambda_{3,8}/2$ (being $\lambda_{3,8}$ the Gell-Mann matrices) and $X$ is the hypercharge operator
associated with the $U(1)_X$ group. In the following subsections we present the field content of the model, with its charges associated with each group on the parentheses, in the form ($SU(3)_C$, $SU(3)_L$, $U(1)_X$).

\subsection{The scalar sector}
\label{subsec:scalars}

The minimal scalar sector for the model is composed by three triplets:
\begin{equation}
\chi=\left(\begin{array}{c}
\chi^{-} \\ \chi^{--} \\ \chi^{0}
\end{array}\right) \sim\left(1,3,+1\right),\quad
\rho=\left(\begin{array}{c}
\rho^{+} \\ \rho^{0} \\ \rho^{++}
\end{array}\right) \sim\left(1,3,-1\right),
\quad
\eta=\left(\begin{array}{c}
\eta^{0} \\ \eta_{1}^{-} \\ \eta_{2}^{+}
\end{array}\right) \sim\left(1,3,0\right),
\end{equation}
where $\chi^0=\frac{|v_\chi|e^{i\theta_\chi}}{\sqrt{2}}\left( 1+\frac{X^0_\chi+i I^0_\chi}{|v_\chi|} \right)$ and
$\psi^0=\frac{|v_\psi|}{\sqrt{2}}\left( 1+\frac{X^0_\psi+i I^0_\psi}{|v_\chi|} \right)$, for $\psi=\,\eta,\,\rho$. We have already rotated away the phases in $v_\eta$ and $v_\rho$ and considered 
them real. 

\subsection{Leptons}
\label{subsec:leptons}

The leptonic sector has three left-handed triplets and six right-handed singlets:
\begin{equation}
\Psi_{aL}=\left(\begin{array}{c}
\nu^\prime_{a} \\ l^{\prime-}_{a}\\ E^{\prime+}_{a}
\end{array}\right) \sim \left(1,3,0\right),
\quad
l^{\prime-}_{aR} \sim \left(1,1,-1\right) \qquad E^{\prime+}_{aR} \sim \left(1,1,1\right),
\label{leptons}
\end{equation}
where the indexes $L$ and $R$ indicate left-handed and right-handed spinors, respectively, $E^\prime_a=E^\prime_e,E^\prime_\mu,E^\prime_\tau$ are new exotic heavy leptons with positive electric charge, and $l^\prime_a=e^\prime,\mu^\prime,\tau^\prime$. Right-handed neutrinos, $\nu_{aR}\sim(1,1,0)$, can be added but, in the present context, they are not important.

The Yukawa Lagrangian in the lepton sector is given by:
\begin{equation}
-\mathcal{L}^{lep}_Y=G^e_{ab}\bar{\Psi}_{aL}l^\prime_{bR}\rho + G^E_{ab}\bar{\Psi}_{aL}E^\prime_{bR}\chi+H.c.,
\label{l1}
\end{equation}
$G^e$ and $G^E$ are arbitrary $3\times3$ matrices in the flavor space, and the mass matrices are given by $M^l=(v_\rho/\sqrt{2})G^e$ and $M^E=(\vert v_\chi \vert/\sqrt{2})e^{i\theta_\chi}G^E$ for the charged and heavy leptons, respectively. We assume for the sake of simplicity that the matrix $G^E$ is diagonal and define $G^E=\vert G^E\vert e^{-i\theta_\chi}$ for the masses of the heavy leptons be real.  Hence, $m_{E_l}=\vert v_\chi\vert  \vert G^E_{ll}\vert /\sqrt{2}$. We have not written the neutrino Dirac and Majorana masses because they are not relevant in the present context. 

The mass eigenstates (unprimed fields) for the charged leptons are related to the symmetry eigenstates (primed fields) through unitary transformations  as $l'_{L,R}=(V_{L,R}^l)^\dagger l_{L,R}$, where
$l_a=(e,\mu,\tau)$.
These $V_{L,R}^{l}$ matrices diagonalize the mass matrix in the following manner:
$V^l_L M^l V^{l\dagger}_R=\hat{M}^l=diag(m_e,m_\mu,m_\tau)$. From this diagonalization we can write $V^l_L M^l M^{l \dagger} V^{l\dagger}_L=(\hat{M}^l)^2$ and $V^l_R M^{l \dagger} M^l V_R^{l \dagger}=(\hat{M}^l)^2$. Solving numerically these equations we obtain one of the possible solutions as
\begin{equation}
V_{L}^{l}=\left(
\begin{array}{ccc}
0.009854 & 0.318482 & -0.947878 \\
0.014571 & -0.947869 & -0.318328 \\
-0.999845 & -0.010674 & -0.013981 \\
\end{array}
\right)
\label{vll}
\end{equation}
\begin{equation}
V_R^l=\left(
\begin{array}{ccc}
0.005014 & 0.002615 & 0.999984 \\
0.007158 & 0.999971 & -0.002650 \\
0.999962 & -0.007171 & -0.004995 \\
\end{array}
\right),
\label{vlr}
\end{equation}
if we use the input for the Yuakawa coupling constants:
\begin{equation}
G^e=\left(
\begin{array}{ccc}
-0.046499 & 0.000374 & 0.000232 \\
-0.000515 & -0.002616 & 0.000014 \\
-0.000657 & -0.000875 & -7.1\times10^{-6} \\
\end{array}
\right)
\label{gel}
\end{equation}
and the observed charged leptons masses. 
To find this solution we have also considered $|v_\rho|=54$ GeV. For the justification of this value see Ref.~\cite{Machado:2013jca}. 

From Eq.~(\ref{l1}) we can write the interactions with the charged scalars:
\begin{eqnarray}
-\mathcal{L}^{lep}_Y&=&\frac{\sqrt2}{v_\rho}\bar{\nu^\prime_L}V^{\dagger l}_L\hat{M}^ll_R\rho^++\frac{\sqrt2}{v_\rho}\bar{E}_LV^{l\dagger}_L\hat{M}^ll_R\rho^{++}\nonumber \\ &+&
\frac{\sqrt2}{\vert v_\chi\vert}e^{-i\theta_\chi}\bar{\nu}_L\hat{M}^EE_R\chi^-+\frac{\sqrt2}{\vert v_\chi\vert}e^{-i\theta_\chi}\bar{l}_LV^l_L\hat{M}^EE_R\chi^{--}+H.c.
\label{l2}
\end{eqnarray}
where $\nu^\prime_L=(\nu^\prime_e\,\nu^\prime_\mu\,\nu^\prime_\tau)^T$. Moreover, the charged scalars have still to be projected over the mass eigenstates denoted by $Y^+_{1,2}$ and $Y^{++}$ (see Appendix~\ref{sec:scalars2}).
Here and below, the vertexes are obtained as usual: $-i\mathcal{L}^{lep}_Y$, and in the lepton case they include the matrices $V^l_{L,R}$ in Eq.~(\ref{vlr}). 

The masses of the heavy leptons are free parameters.
In order to have massive neutrinos, right-handed neutrinos can be added. A Dirac mass for neutrinos is obtained which is proportional to $ v_\rho$ or we can add a scalar sextet $\sim(1,6^*,0)$ coupled to $\overline{(\Psi_L)^c}\Psi_L$ to obtain a Majorana mass term for the active neutrinos. Moreover, if right-handed neutrinos have a Majorana mass term, the model implements a symmetric $6\times6$ neutrino mass matrix. We will address the neutrino masses elsewhere, showing that it is possible to obtain a realistic PMNS matrix, but at present we ignore the neutrino masses. 

\subsection{Quarks}
\label{subsec:quarks}

In the quark sector there are two antitriplets and one triplet, all left-handed, besides the corresponding right-handed singlets:
\begin{equation}
Q_{mL}=\left(\begin{array}{c}
d_{m} \\ -u_{m} \\ j_{m}
\end{array}\right)_L \sim \left(3,3^{*},-1/3\right) ,\qquad
Q_{3L}=\left(\begin{array}{c}
u_{3} \\ d_{3} \\ J
\end{array}\right)_L \sim \left(3,3,2/3\right)
\end{equation}

\begin{equation}
u_{\alpha R} \sim \left(3,1,2/3\right) ,\quad d_{\alpha R} \sim \left(3,1,-1/3\right) ,\quad j_{mR} \sim \left(3,1,-4/3\right),
\quad J_{R} \sim \left(3,1,5/3\right)
\end{equation}
where $m=1,2$ e $\alpha=1,2,3$. The $j_m$ exotic quarks have electric charge -4/3 and the $J$ exotic quark has electric charge 5/3 in units of $\vert e \vert$.

The Yukawa interactions between quarks and scalars are given by
\begin{eqnarray} 
-\mathcal{L}^q_Y &=& \bar{Q}_{mL} \left[ G_{m\alpha} U^{'}_{\alpha R} \rho^*+\tilde{G}_{m\alpha}D^{'}_{\alpha R} \eta^* \right]+
\bar{Q}_{3L} \left[ F_{3\alpha}U^{'}_{\alpha R} \eta + \tilde{F}_{3\alpha}D^{'}_{\alpha R} \rho \right] \nonumber \\&  +&
\bar{Q}_{mL}G'_{mi}j_{iR}\chi^* + \bar{Q}_{3L}g_J J_R \chi +H.c.,
\label{q1}
\end{eqnarray}
where we omitted the sum in $m$, $i$ and $\alpha$, $U^{'}_{\alpha R}=(u^\prime\,c^\prime\,t^\prime)_R$ and $D^{'}_{\alpha R}~=~
(d^\prime\,s^\prime\,b^\prime)_R$. $G_{m\alpha}$, $\tilde{G}_{m\alpha}$, $F_{3\alpha}$, $\tilde{F}_{3\alpha}$, $G'_{mi}$ and $g_J$ are the coupling constants.

From Eq.~(\ref{q1}), we obtain that the exotic quarks have the following interactions with the charged scalars
\begin{eqnarray}
-\mathcal{L}_j&=&
\bar{\tilde{j}}_L[\mathcal{O}^u
V^U _R\, U_R +\mathcal{O}^dV^D_R\, D_R] +\frac{\sqrt2}{\vert v_\chi\vert}\,\bar{D}_LV^D_L \left(\begin{array}{ccc}
m_{j_1}\chi^+ & 0 & 0\\
0 & m_{j_2}\chi^+&0\\
0&0&m_J\chi^{--}
\end{array}\right)\tilde{j}_R
\nonumber \\&+&
\frac{\sqrt2}{\vert v_\chi\vert}\,\bar{U}_L \,V^U_L \left(\begin{array}{ccc}
m_{j_1}\chi^{++}& 0 & 0\\
0 & m_{j_2}\chi^{++}&0\\
0&0&m_J\chi^{-}
\end{array}\right)\tilde{j}_R+
H.c.
\label{jjJ}
\end{eqnarray}
where $\tilde{j}=(j_1\,j_2\,J)^T$, $U_{L,R}=(u\,c\,t)^T_{L,R}$ and $D_{L,R}=(d\,s\,b)^T_{L,R}$ denote the mass eigenstates.
We have defined the matrices 
\begin{equation}
\mathcal{O}^u=\left(\begin{array}{ccc}
G_{11}\rho^{--}&G_{12}\rho^{--}&G_{13}\rho^{--}\\
G_{21}\rho^{--}&G_{22}\rho^{--}&G_{23}\rho^{--}\\
F_{31}\eta^+_2 &F_{32}\eta^+_2 &F_{33}\eta^+_2 \\
\end{array}\right),\;\;\mathcal{O}^d=
\left(\begin{array}{ccc}
\tilde{G}_{11}\eta^-_2&\tilde{G}_{12}\eta^-_2 &\tilde{G}_{13}\eta^-_2 \\
\tilde{G}_{21}\eta^-_2 &\tilde{G}_{22}\eta^-_2 &\tilde{G}_{23}\eta^-_2 \\
\tilde{F}_{31}\rho^{++}  &\tilde{F}_{32}\rho^{++}  &\tilde{F}_{33}\rho^{++}  \\
\end{array}\right).
\label{J}
\end{equation}

In Eq.~(\ref{jjJ}) we have assumed that the mass matrix in the $j_1,j_2$ sector is diagonal, i.e.,  $G'_{12}=G'_{21}=0$. In this case $G_{ii}=\vert G_{ii}\vert e^{i\theta_\chi}$ and $g_J=\vert g_J\vert e^{i\theta_\chi}$. After absorbing the $\theta_\chi$ phase in the masses we have $\vert g_J\vert=m_J\sqrt{2}/\vert v_\chi\vert$ and $\vert G_{ii}\vert=m_{j_i}\sqrt{2}/\vert v_\chi\vert$.  
We have also used the fact that if $U^{'}_{L,R}$ and $D^{'}_{L,R}$ denote the symmetry eigenstates and $U_{L,R}$ and $D_{L,R}$ the mass eigenstates, they are related by unitary matrices as follows: $U^{'}_{L,R}=\left( V_{L,R}^U\right)^\dagger U_{L,R}$ and $D^{'}_{L,R}=\left( V_{L,R}^D\right)^\dagger D_{L,R}$ in such a way that
$V_{L}^{U} M^u V_{R}^{U\dagger}=\hat{M}^u=diag(m_u,m_c,m_t)$ and
$V_{L}^{D} M^d V_{R}^{D\dagger}=\hat{M}^d=diag(m_d,m_s,m_b)$.

In terms of the mass eigenstates we can write the Yukawa interactions in Eqs.~(\ref{jjJ}) and (\ref{J}) as in Appendix~\ref{sec:quarkscalars}, where the charged scalar has already been projected on the physical $Y^-_2,Y^{--}$. In this appendix we wrote only the interactions which appear in the EDM diagrams. 

Using as input the observed quark masses and the mixing matrix in the quark sector, $V_{CKM}=V_L^U
V_L^{D\dagger}$~\cite{Agashe:2014kda}, the numerical values of the matrices $V^{U,D}_{L,R}$ were found to be \cite{Machado:2013jca}
\begin{eqnarray}
&& V^U_L=\left(\begin{array}{ccc}
	-0.00032 & 0.00433 & 0.99999 \\
	0.07163 & -0.99742 & 0.00434 \\
	-0.99743 & -0.07163 & -0.00001 \\
\end{array}\right),\nonumber \\&&
 V^D_L\!\!=\!\!\left(\begin{array}{ccc}
	0.004175 & -0.209965 & 0.97761 \\
	0.03341 & -0.977145 & -0.209995 \\
	-0.999525 & -0.03052 & -0.004165 \\
\end{array}\right)
\label{vudl331}.
\end{eqnarray}

In the same way we obtain the $V^{U,D}_R$ matrices:
\begin{eqnarray}
&& V^U_R=\left(\begin{array}{ccc}
	-0.4544 & 0.13857 & 0.87996 \\
	0.82278 & -0.31329 & 0.47421 \\
	-0.34139 & -0.93949 & -0.02834 \\
\end{array}\right),\nonumber  \\&&
 V^D_R\!\!=\!\!\left(\begin{array}{ccc}
	-0.0001815 & -0.325355 & 0.94559 \\
	0.005976 & -0.945575 & -0.325345 \\
	-0.999982 & -0.00559 & -0.002115 \\
\end{array}\right).
\label{vudr331}
\end{eqnarray}

It should be noted that the product $V^U_L V^{D\dagger}_L$ of the matrices above correspond to the CKM matrix when the modulus is considered. The known quark masses depend on both $v_\eta$ and $v_\rho$. The values of the matrices $V^{U,D}_{L,R}$ were obtained by using $v_\rho=54$ GeV and $v_\eta=240$ GeV. The matrices given in Eqs.~(\ref{vudl331}) and (\ref{vudr331}) give the correct quark masses (at the $Z$-pole given in Ref.~\cite{Machado:2013jca}) and the CKM matrix if the Yukawa couplings are:  $G_{11}=1.08,G_{12}=2.97,G_{13}=0.09,G_{21}=0.0681,
G_{22}=0.2169,G_{23}=0.1\times10^{-2}$,
$F_{31}=9\times10^{-6},F_{32}=6\times10^{-6},F_{33}=1.2\times10^{-5}$, $\tilde{G}_{11}=0.0119,
\tilde{G}_{12}=6\times10^{-5},\tilde{G}_{13}=2.3\times10^{-5},\tilde{G}_{21}=(3.2 - 6.62)\times10^{-4},\tilde{G}_{22}=
2.13\times10^{-4},\tilde{G}_{23}=7\times10^{-5}$, $\tilde{F}_{31}=2.2\times10^{-4},
\tilde{F}_{32}=1.95\times10^{-4},\tilde{F}_{33}=1.312\times10^{-4}$. All these couplings should be multiplied by $\sqrt{2}$; it is a conversion factor from the notation used in \cite{Machado:2013jca} to our notation. We also took the central values of the matrices $V^D_{L,R}$ presented in this reference for our calculations.

\section{The EDM in this model}
\label{sec:edmft}

In the framework of quantum field theory (QFT) the EDM of a fermion is described by an effective Lagrangian
\begin{equation}
\mathcal{L}_{EDM}=-i\sum_{f}\frac{d}{2}\bar{f}\sigma^{\mu\nu}\gamma_{5}fF_{\mu\nu}
\label{eq1}
\end{equation}
where $d$ is the magnitude of the EDM, $f$ is the fermion wave function and $F_{\mu\nu}$ is the electromagnetic tensor. This Lagrangian
gives rise to the vertex
\begin{equation}
\label{vertice_MDE}
\Gamma^{\mu}=id\sigma^{\mu\nu}q_{\nu}\gamma_{5}
%\label{eq2}
\end{equation}
where $q_{\nu}$ is the photon's momentum.

Since the EDM is an electromagnetic property of a particle, its Lagrangian depends on the interaction between the particle and the electromagnetic field. To find the EDM, one must consider all the diagrams for a vertex between the particle and a photon. The sum of the
amplitudes will be proportional to
\begin{equation}
\label{vertice_generico_MDM_eletron}
\Gamma^{\mu}\left(q\right)=F_{1}\left(q^{2}\right)\gamma^{\mu}+\cdots+F_{3}\left(q^{2}\right)\sigma^{\mu\nu}\gamma_{5}q_{\nu}
%\label{eq3}
\end{equation}
Comparing with Eq. \ref{vertice_MDE}, we can see that $d=\textrm{Im}[F_3(0)]$.  

\subsection{The electron EDM}
\label{subsec:edmleptons}

\begin{figure}
\includegraphics[width=0.75\textwidth]{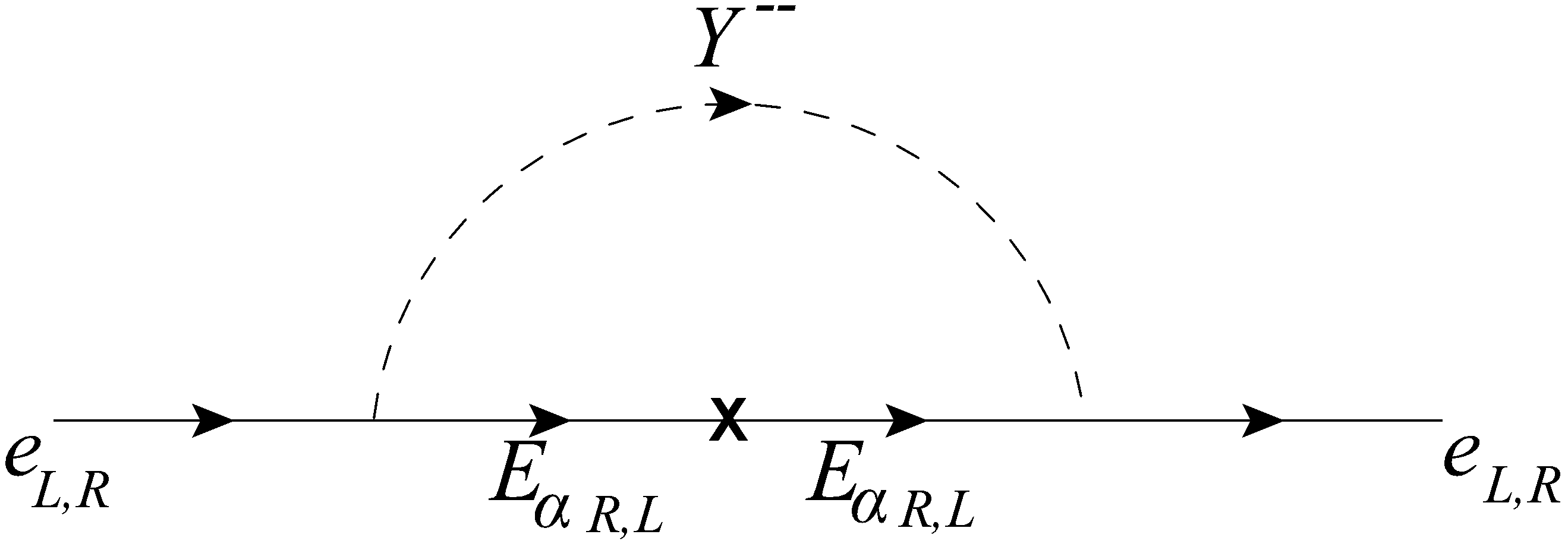}
\caption{Diagrams contributing to the electron EDM. It should be considered the case where the photon line is connected
to the scalar line and the case where it is connected to the fermion line. Also, all the left-right combinations and all
the exotic lepton possibilities ($\alpha=e,\,\mu\,\tau$) should be considered.}
\label{MDE_eletron_todos}
\end{figure}

Considering the diagrams like that given in Fig.~\ref{MDE_eletron_todos} we find the following expression for the electron EDM contributions at the one-loop level. Assuming that the only source of $CP$ violation are the $e$-$E_l$-$Y$ vertexes in Eq.~(\ref{l2}) with $\rho^{--}$ and $\chi^{--}$ projected on $Y^{--}$ as is shown in Eq.~(\ref{a1}), the electron EDM is given by:
\begin{eqnarray} 
\left.\frac{d_e}{e\cdot \textrm{cm}}\right|_Y&=&\left\lbrace \textrm{Im} \Big[(V_{E_Ll_R})_{11}(V_{l_LE_R})_{11}\Big]-\textrm{Im}\Big[(V^\dagger_{l_LE_R})_{11}(V^\dagger_{E_Ll_R})_{11}\Big]\right\rbrace\Big[I^{eE_eY}_1+2 I^{eE_eY}_2\Big]
\nonumber \\&+ &\left\lbrace \textrm{Im}\Big[(V_{E_Ll_R})_{21}(V_{l_LE_R})_{12}\Big]-\textrm{Im}\Big[(V^\dagger_{l_LE_R})_{21}(V^\dagger_{E_Ll_R})_{12}\Big]\right\rbrace \Big[I^{eE_\mu Y}_1+2 I^{eE_\mu Y}_2\Big]
\nonumber \\& +& \left\lbrace \textrm{Im}\Big[(V_{E_Ll_R})_{31}(V_{l_LE_R})_{13}\Big]-\textrm{Im}\Big[(V^\dagger_{l_LE_R})_{31}(V^\dagger_{E_Ll_R})_{13}\Big]\right\rbrace \Big[I^{eE_\tau Y}_1+2 I^{eE_\tau Y}_2\Big]
\nonumber \\&=&  -(197 \times 10^{-16}\;GeV) \frac{2\sin(2\theta_\chi)}{\vert v_\rho\vert
 \left(1+
\frac{\vert v_\chi\vert^2}{
\vert v_\rho\vert^2}\right)
}
\nonumber \\ &.& 
\left\{
m_{E_e}\left[ (V_L^l)_{11}\sum_i G^e_{1i}(V_R^l)_{i1} \right]\left[I^{eE_e Y}_1+2 
I^{eE_e Y}_2\right] 
+m_{E_\mu}\left[(V_L^l)_{21}\sum_i G^e_{2i}(V_R^l)_{i1} \right]\right.\nonumber \\& .& \left.\left[I^{eE_\mu Y}+2 
I^{eE_\mu Y} \right]
+ m_{E_\tau}\left[(V_L^l)_{31}\sum_i G^e_{3i}(V_R^l)_{i1} \right]\left[I^{eE_\tau Y}_1+2 
I^{eE_\tau Y}_2\right] \right\}.
\label{eq:mde_eletron}
\end{eqnarray}
where $Y$ denotes $Y^{++}$.
The elements of the matrices $V_{E_Ll_R}$ and $V_{l_LE_R}$ of the above equation can be found in Appendix~\ref{sec:leptonscalars}. The factors $I^{eE_lY}_1$ and $I^{eE_lY}_2$ are given by

\begin{equation} \label{eq:I1}
I^{el}_1\equiv I_1(m_{E_l},m_e,m_Y)=-\frac{m_{E_l}}{4(4\pi)^2}\int\limits_{0}^{1} dz
\frac{1+z}{(m^2_{E_l}-zm^2_e) (1-z)+m_Y^2 z},
\end{equation}
and
\begin{equation} \label{eq:I2}
I^{el}_2\equiv  I_2(m_{E_l},m_e,m_Y)=-\frac{m_{E_l}}{4(4\pi)^2}\int\limits_{0}^{1} dz
\frac{z}{[m^2_{E_l}-(1-z)m^2_e] z+m_Y^2(1-z)},
\end{equation}
where $m_{E_l}$ $(l=e,\mu,\tau)$ and $m_e$  denote the mass of the heavy lepton and the electron mass respectively. Also, $m_Y$ is the mass of the scalar in the diagram, which in this case is $Y^{++}$.
 % % % % GEORGE

\begin{figure}
\includegraphics[width=.75\textwidth,origin=c,angle=0]{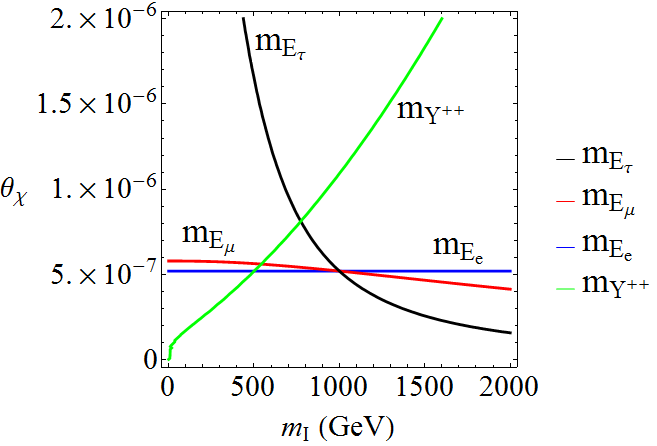}
%./EDM_theta_mYPP
\caption{
Allowed values for the exotic particles masses shown in the figure and $\theta_\chi$ from
the electron EDM. The regions below each line show the allowed values
for $\theta_\chi$ and the heavy lepton masses $m_{E_l}$ that satisfy $\vert d_e\vert_Y<8.7\times 10^{-29}$ e cm; see
Eq.~(\ref{eq:mde_eletron}). Each line corresponds to the mass of a
different particle (as shown in the legend). For a given line, the
electron EDM is evaluated considering the value presented in the lower
axis for the corresponding mass, while the other masses have their values fixed (for more information see the text).}
\label{edme}
\end{figure}

Using Eq.~(\ref{eq:mde_eletron}) and considering that it respects the actual experimental limit \cite{Baron:2013eja} ($|d_e|_Y<|d_e|_{\textrm{exp}}=8.7 \times 10^{-29}\,e\cdot \textrm{cm}$) we obtain the graph in Fig.~\ref{edme}. The regions below each line indicate the values for $\theta_\chi$ and $m_I$ ($m_I$ being the mass of the heavy particle in the internal line) where our theoretical prediction is in agreement with the experimental results. Each line corresponds to the mass of a different particle (as shown in the legend). For a given line, the electron EDM is evaluated considering the value presented in the lower axis for the corresponding mass, for the other masses the values are
taken to be (in GeV):  $m_{Y^{++}}=500$, $m_{E_e}=1000$, $m_{E_\mu}=1000$ and $m_{E_\tau}=1000$. We also considered $|v_\chi|=2000$ GeV and $\vert v_\rho\vert=54$ GeV. The values of the matrix entries $V^l_{L,R}$ and of the $G^e$ Yukawa couplings are those given in Eqs.~(\ref{vll}), (\ref{vlr}) and (\ref{gel}), respectively. Notice that the projection over the mass eigenstate $Y^{--}$ implies the factor  
\begin{equation} 
\frac{1}{ \vert v_\rho\vert}\,
\frac{1}{
1+\frac{\vert v_\chi\vert^2}{\vert v_\rho\vert^2}
}\approx \frac{\vert v_\rho\vert}{\vert v_\chi\vert^2}\sim \;1.34\times10^{-5}\,\textrm{GeV}^{-1}
\label{sup} 
\end{equation} 
 It should be noted that our theoretical prediction only allows small values for $\theta_\chi$, being of order $10^{-6}$ to $10^{-7}$, except in the case where the $E_\tau$ mass is small or the $Y^{--}$ mass is large.

\subsection{The neutron EDM}
\label{subsec:neutronedm}

\begin{figure}
\includegraphics[width=0.75\textwidth]{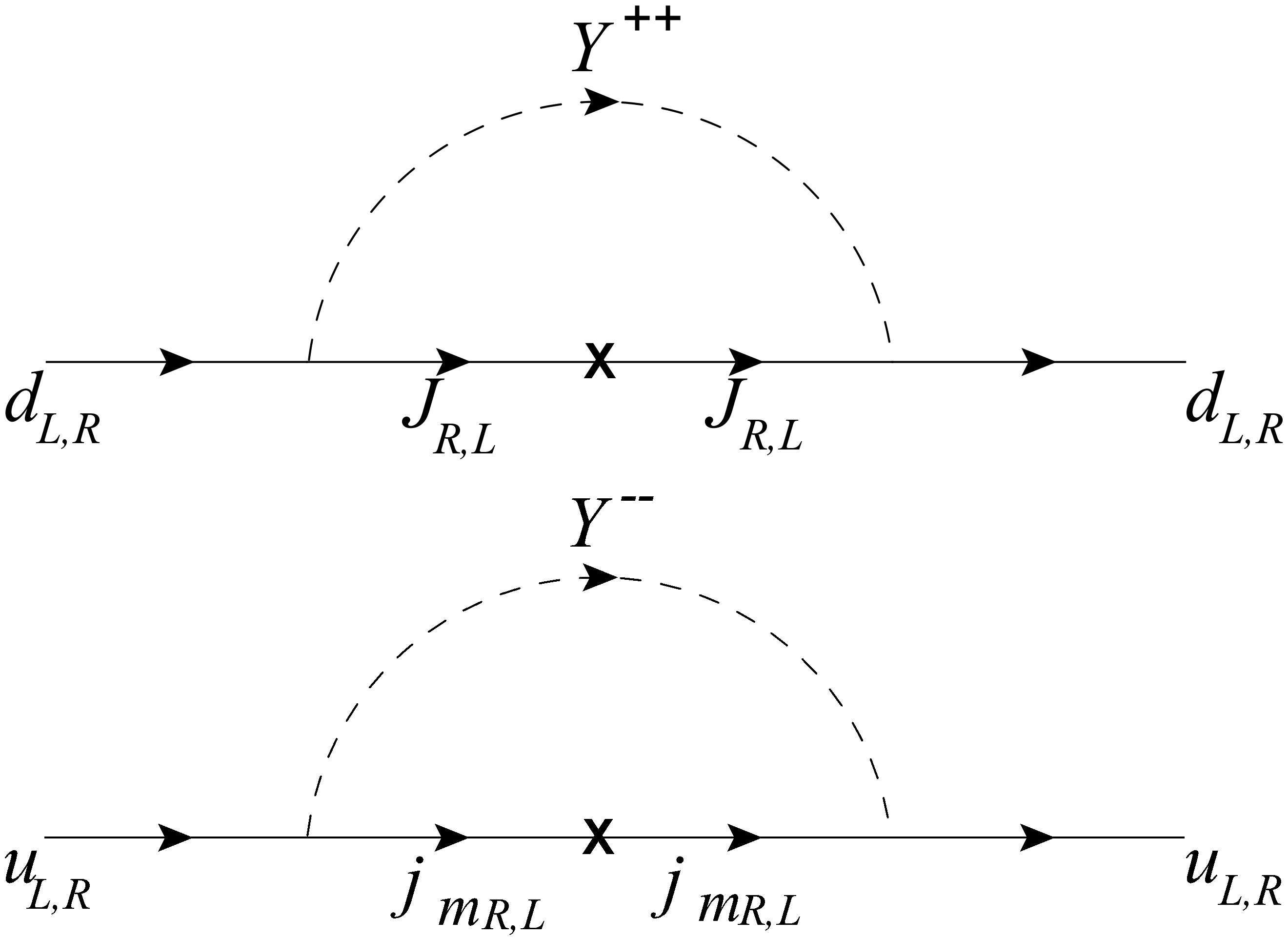}
\caption{Diagrams contributing to the neutron EDM. For each diagram the case should be considered where the photon line is connected to the scalar line and where it is connected to the fermion line. Also, all the left-right combinations and all	the exotic quark possibilities for the $u$-quark diagram ($m=1,\,2$) should be considered.}
\label{fig:edm1}
\end{figure}

As in the case of charged leptons we will assume here that the only source of $CP$ violation is the phase $\theta_\chi$. Considering the diagrams given in Fig.~\ref{fig:edm1} we find an expression for the neutron EDM in the 3-3-1 model with heavy leptons. For each diagram we calculate the contribution to the EDM given by each quark, with the total EDM of the neutron given by:
\begin{equation} \label{eq:mde_331}
d_n\vert_Y=\left(\frac{4}{3}d_d-\frac{1}{3}d_u\right)_Y
\end{equation}
where
\begin{eqnarray}
%\begin{split}
\left.\frac{d_d}{e\cdot\textrm{cm}}\right\vert_Y&=&\left\lbrace \textrm{Im}\Big[(K_{J_LD_R})_{31}(K_{D_LJ_R})_{13}\Big]-\textrm{Im}\Big[(K^\dagger_{D_LJ_R})_{31}(K^\dagger_{J_LD_R})_{13}\Big]\right\rbrace \Big[I^{dJY}_1 +2 I^{dJY}_2) \Big]
\nonumber \\& =&
-(197 \times 10^{-16}\;\text{GeV})\left[ \sin(2\theta_\chi) m_J \frac{2\sqrt{2}|v_\rho|}{|v_\rho|^2+|v_\chi|^2}(V^D_L)_{13} \sum_{k} (V^D_R)_{1k} \tilde{F}_{3k} \right]\Big[I^{dJY}_1+2 I^{dJY}_2\Big]
%\end{split}
\label{edmn1}
\end{eqnarray}
where $Y$ denotes $Y^{++}$. We have used the definition of the matrices in Eqs.~(\ref{ec2}) and (\ref{ec8}).

Similarly, considering the figures involving the $u$ quark in Fig.~\ref{fig:edm1}
\begin{eqnarray}
\left.\frac{d_u}{e\cdot\textrm{cm}}\right\vert_Y&=&\left\lbrace \textrm{Im}\Big[ (K_{j_LU_R})_{11}(K_{U_Lj_R})_{11}\Big]-\textrm{Im}\Big[(K^\dagger_{U_Lj_R})_{11}(K^\dagger_{j_LU_R})_{11} \Big] \right\rbrace \Big[ I^{uj_1Y}_1+2 I^{uj_1Y}_2 \Big]
\nonumber \\& 
+& \left\lbrace \textrm{Im}\Big[ (K_{j_LU_R})_{21}(K_{U_Lj_R})_{12}\Big]-\textrm{Im}\Big[(K^\dagger_{U_Lj_R})_{21}(K^\dagger_{j_LU_R})_{12} \Big] \right\rbrace \Big[ I^{uj_2Y}_1+2 I^{uj_2Y}_2 \Big]
\nonumber\\& =& (197 \times 10^{-16}\;\text{GeV}) 
\left[ \sin(2\theta_\chi) m_{j1} \frac{2\sqrt{2}|v_\rho|}{|v_\rho|^2+|v_\chi|^2}(V^U_L)_{11} \sum_{k} (V^U_R)_{1k} G_{1k} \right] \Big[ I^{uj_1Y}_1+2 I^{uj_1Y}_2 \Big]
\nonumber \\& 
+&(197 \times 10^{-16}\;\text{GeV}) \left[\sin(2\theta_\chi) m_{j2} \frac{2\sqrt{2}|v_\rho|}{|v_\rho|^2+|v_\chi|^2}(V^U_L)_{12} \sum_{k} (V^U_R)_{1k} G_{2k}\right] \Big[ I^{uj_2Y}_1+2 I^{uj_2Y}_2 \Big],
\label{edmn2}
\end{eqnarray}
and the integrals $I^{dj_1Y}_{1,2}$ are given in Eqs.~(\ref{eq:I1}) and (\ref{eq:I2}), respectively, making $m_e\to m_d,m_{E_l}\to m_{j_1}$ and $m_Y\to m_{Y^+_2}$ and for $I^{dJY}_{1,2}$ making $m_e\to m_u,m_{E_l}\to m_J$ while $m_Y$ is the same. We used the matrices defined in Eqs.~(\ref{ec6}) and (\ref{ec12}). Notice that only the exotic quarks with charge 5/3 contribute to the $d$-quark EDM and only those with electric charge $-4/3$ do for the $u$-quark EDM.

On the equations above we have considered the $V^U_{L,R}$ and $V^D_{L,R}$ matrices to be real, that is because we considered the numerical results presented in \cite{Machado:2013jca} for such matrices and for the Yukawa couplings [see Eqs.~(\ref{vudl331}) and (\ref{vudr331})].

% % % % % % % % % % % % % % % % %george
Using Eq. (\ref{eq:mde_331}) and considering that it respects the actual experimental limit \cite{Baker:2006} ($\vert d_Y\vert <\vert d_n\vert_{\textrm{exp}}=2.9 \times 10^{-26}\,e\cdot \textrm{cm}$) we obtain the graph in Fig. \ref{fig:edmneutron1}. Similar to the electron case, the regions below each line indicate the values for $\theta_\chi$ and $m_I$ ($m_I$ being the mass of the exotic particle in the internal line) where our theoretical prediction is in agreement with the experimental results. Each line corresponds to the mass of a different particle (as shown in the legend). For a given line, the neutron EDM is evaluated considering the value presented in the lower axis for the corresponding mass, for the other masses the values are taken to be (in GeV): $m_{Y_2^+}=200$, $m_{Y^{++}}=500$, $m_J=1000$, $m_{j_1}=1000$ and $m_{j_2}=1000$. We also considered $v_\eta=240$ GeV, $v_\rho=54$ GeV, and $|v_\chi|=2000$ GeV. The values for the Yukawa couplings used in Eq.~(\ref{J}) are those below Eq.~(\ref{vudr331}).

The graphs indicates that smaller masses for the exotic quark $J$ and large masses for the exotic scalar $Y^{++}$ are favored, while the EDM seems unaffected by changes in the masses of the other exotic quarks or $Y^+_2$. For the neutron we also find that $\theta_\chi$ should have a small value, of order $10^{-1}$. However, the results for the electron yields even smaller limits for $\theta_\chi$, leaving room for a greater range of possible values for $m_J$ and $m_{Y_2^+}$.

\section{Conclusions}
\label{sec:con}

The electron EDM imposes a strong constraint in the new mechanism of $C\!P$ violation.  Both the experimental upper limit and the SM prediction are lower than the neutron EDM. Moreover it is not sensitive to QCD corrections, at least at the 1-loop level. 
In the framework of the 3-3-1 models, the electron EDM was calculated in Refs.~\cite{Liu:1993gy,Montero:1998yw}. However, at that time we knew nothing about either the unitary matrices in the lepton sector, $V^l_{L,R}$, or $V^{U,D}_{L,R}$ in the quark sector. Notwithstanding, after the results reported from Ref.~\cite{Machado:2013jca} it is possible to make more realistic calculations of the EDM since now the number of free parameters is lower than before. In fact, once the values of $\vert v_\rho\vert$ and $\vert v_\eta\vert$ are obtained, the quark masses and the CKM matrix determine, not necessarily univocally, the unitary matrices in the quark sector. The same happens in the lepton sector as is shown in Sec.~\ref{subsec:leptons}. At this level, the unknown parameters are the phase $\theta_\chi$, the masses of the scalars (although one of the neutral ones has to have a mass of the order of 125 GeV), the orthogonal matrix which diagonalize the mass matrix of the \textit{CP} even neutral scalars, and the masses of the exotic quarks. 

\begin{figure}[tbp]
\centering % \begin{center}/\end{center} takes some additional vertical space
\includegraphics[width=.75\textwidth,origin=c,angle=0]{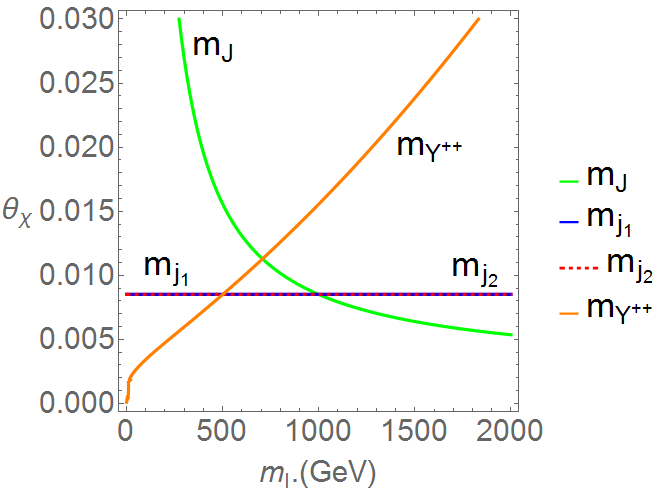}
%./EDM_neutron_mY2
\hfill
\caption{
Allowed values for the exotic particles masses shown in the figure and $\theta_\chi$ from
the neutron EDM. The regions below each line show the allowed values
for $\theta_\chi$ and exotic particle masses that satisfy $|d_n|_Y<2.9\times 10^{-26}$ e  cm; see
Eqs.~(\ref{eq:mde_331}) - (\ref{edmn2}). Each line corresponds to the
mass of a different particle (as shown in the legend). For a given
line, the neutron EDM is evaluated considering the value presented in
the lower axis for the corresponding mass, while the other masses have their values fixed (for more information see the text).}
\label{fig:edmneutron1}
\end{figure}

From the calculation of the EDM of the neutron and the electron at 1-loop order, we were able to set  lower limits on the masses of the $Y_2^+$ and $Y^{++}$ scalars, which are compatible with the search of these sorts of fields at the LHC and Tevatron~\cite{Davey:2014tka},  and on the masses of the exotic fermions, depending on the value of $\theta_\chi$, and we have also a good indication that this phase should be below $10^{-6}$. 
From the graph in Fig. \ref{edme} we see that as the mass of
$Y^{++}$ goes up the electron EDM decreases, while the inverse happens
for the mass of $E_{\tau}$. In the case of the neutron EDM, from Fig.
\ref{fig:edmneutron1}, we see also that the increase of the mass of the scalar
$Y^{++}$ decreases the EDM, and the decrease of $m_{J}$
(the mass of the exotic quark $J$) also decreases the EDM. Analyzing
Eqs.~(\ref{eq:I1}) and (\ref{eq:I1}) it is clear that the increase of the masses of the
exotic scalars will decrease the EDM, since these masses appear in the
denominator. As for the decrease of the EDM from the decrease of the
masses of the exotic fermions it can be explained from the fact that
 Eqs.~(\ref{eq:I1}) and (\ref{eq:I1}) are proportional to those masses. However, this is not
the only thing to be taken into account, because from Fig. \ref{fig:edmneutron1} we
see that the decrease of $m_{E_\mu}$ and $m_J$ increases the EDM. This
effect can be explained from the signs of the coupling constants and
elements of the fermion diagonalization matrices, which can lead to
cancellations among the many diagrams involved in the final result.

It seems that in this model we have a situation similar with that in supersymmetric theories in which the EDM's are larger than the SM prediction and are appropriately suppressed only by the phases. This is the so-called SUSY $C\!P$-problem. See Refs.~\cite{Pospelov:2005pr,Ritz:2009zz} and references therein. We stress again that we have considered only the soft $C\!P$ violation present in the model. In fact, it has other $C\!P$ hard violating sources. Beside the phase $\delta$ in the CKM matrix, the matrices $V^{U,D,l}_{L,R}$ are also complex with, in principle, six arbitrary phases. 
In the SM, the contribution of the CKM matrix $\delta$ to $d_{e,n}$ is negligible at the 1-loop level in pure weak amplitudes, but this is not necessarily the case for the phases in the matrices $V^{U,D,l}_{L,R,l}$.
For instance, if the matrix $V^l_L$ is complex the electron EDM in Eq.~(\ref{eq:mde_eletron}) will be proportional to $2\sin(2\theta_\chi\pm\theta_{V^l_L}\mp\theta_{V^l_R})$, where $\theta_{V^l_L},\theta_{V^l_R}$ denote the extra phases from the respective matrices. In this case, all phases may be naturally of $O(1)$ while the sum is small $\sim10^{-6}$. 

The contributions of these phases in the framework of the minimal 3-3-1 model were done in Ref.~\cite{Liu:1993gy}. It is, of course, important to take into account these extra phases, but it is beyond the scope of the present work. We recall that even the right-handed matrices $V^{U,D}_R$ survive in the neutral scalar sector which has flavor changing neutral currents as it was shown in Ref.~\cite{Machado:2013jca}. It is possible that three of the phases in $V^D_L$ can be absorbed in the exotic quarks $J,j_1$ and $j_2$, but there is no more freedom to absorb the phases in $V^U_L$. Notwithstanding, these phases will appear in the vertexes shown in Sec.~\ref{sec:quarkscalars}.

\acknowledgements

G. D. C. would like to thank CAPES and CNPq for the financial support and V. P. would like to thank CNPq for partial support.

\newpage
\appendix

\section{The scalar sector}
\label{sec:scalars2}

The most general potential, invariant under CP transformations, for the scalars is
\begin{eqnarray}
\label{potencial_escalar}
V(\chi,\eta,\rho)&=& \sum_{i} \mu^2_{i} \phi^\dagger_i \phi_i
+\sum_{i=1,2,3}a_i(\phi^{\dagger}_i\phi_i)^2 +\sum_{m=4,5,6,i>j} a_m(\phi_i^\dagger \phi_i)(\phi_j^\dagger \phi_j)
\nonumber \\&+&
\sum_{n=7,8,9;i>j}a_n (\phi^{\dagger}_i\phi_j)(\phi^\dagger_j\phi_i) +( \alpha \,\epsilon_{ijk}\chi_{i}\rho_{j}\eta_{k}+H.c.) ,
\label{potential}
\end{eqnarray}
where we have used $\phi_1=\chi,\phi_2=\eta$ and $\phi_3=\rho$, except in the trilinear term.

Taking the derivatives of Eq.~(\ref{potencial_escalar}) with respect to the vacua and setting these to zero we are able to
find expressions for $\mu^2_\chi$, $\mu^2_\eta$ and $\mu^2_\rho$. Also, from these derivatives, we can find that
$\alpha=|\alpha|e^{-i \theta_\chi}$. Using this we can find the mass matrices and, therefore, the following mass eigenstates:

Double charge scalars:
\begin{eqnarray}
&&\left(\begin{array}{c}
\rho^{++} \\ \chi^{++}
\end{array}\right)=\frac{1}{\sqrt{1+\frac{|v_\chi|^2}{|v_\rho|^2}}}
\left(\begin{array}{cc}
1 & \frac{|v_\chi|}{|v_\rho|}e^{-i\theta_\chi} \\ -\frac{|v_\chi|}{|v_\rho|}e^{i\theta_\chi} & 1
\end{array}\right)
\left(\begin{array}{c}
G^{++} \\ Y^{++}
\end{array}\right)
\nonumber \\&&
m^2_{G^{++}}=0,\quad
m^2_{Y^{++}}=A\left(\frac{1}{|v_\rho|^2}+\frac{1}{|v_\chi|^2}\right)+\frac{a_8}{2}\left(|v_\chi|^2+|v_\rho|^2\right),
\label{a1}
\end{eqnarray}
where $ A= |v_\chi||v_\eta||v_\rho||\alpha|/\sqrt{2}$.

First pair of single charge scalars:
\begin{eqnarray}
&&\left(\begin{array}{c}
\eta_1^+ \\ \rho^+
\end{array}\right)=\frac{1}{\sqrt{1+\frac{|v_\rho|^2}{|v_\eta|^2}}}
\left(\begin{array}{cc}
1 & \frac{|v_\rho|}{|v_\eta|} \\ -\frac{|v_\rho|}{|v_\eta|} & 1
\end{array}\right)
\left(\begin{array}{c}
G_1^+ \\ Y_1^+
\end{array}\right),
\nonumber \\&&
m^2_{G_1^+}=0,\quad
m^2_{Y_1^+}=A\left(\frac{1}{|v_\rho|^2}+\frac{1}{|v_\eta|^2}\right)+\frac{a_9}{2}\left(|v_\eta|^2+|v_\rho|^2\right).
\label{a2}
\end{eqnarray}

Second pair of single charge scalars:
\begin{eqnarray}
&&\left(\begin{array}{c}
\eta_2^+ \\ \chi^+
\end{array}\right)=\frac{1}{\sqrt{1+\frac{|v_\chi|^2}{|v_\eta|^2}}}
\left(\begin{array}{cc}
1 & \frac{|v_\chi|}{|v_\eta|}e^{i\theta_\chi} \\ -\frac{|v_\chi|}{|v_\eta|}e^{-i\theta_\chi} & 1
\end{array}\right)
\left(\begin{array}{c}
G_2^+ \\ Y_2^+
\end{array}\right)
\nonumber \\&&
m^2_{G_2^+}=0,\quad
m^{2}_{Y_2^+}=A\left(\frac{1}{|v_\chi|^2}+\frac{1}{|v_\eta|^2}\right)+\frac{a_7}{2}\left(|v_\eta|^2+|v_\chi|^2\right).
\label{a3}
\end{eqnarray}

Neutral CP-odd scalars:
\begin{eqnarray}
&&\left(\begin{array}{c}
I^0_\eta \\ I^0_\rho \\ I^0_\chi
\end{array}\right)=
\left(\begin{array}{ccc}
\frac{N_a}{|v_\chi|} & - \frac{N_b|v_\eta||v_\chi|}{|v_\rho|(|v_\eta|^2+|v_\chi|^2)} & \frac{N_c}{|v_\eta|} \\
0 & \frac{N_b}{|v_\chi|} & \frac{N_c}{|v_\rho|} \\
-\frac{N_a}{|v_\eta|} & - \frac{N_b|v_\eta|^2}{|v_\rho|(|v_\eta|^2+|v_\chi|^2)} & \frac{N_c}{|v_\chi|}
\end{array}\right)
\left(\begin{array}{c}
G^0_1 \\ G^0_2 \\ h^0
\end{array}\right)
\nonumber \\&&
m^2_{G_1^0}=m^2_{G_2^0}=0,\quad
m^2_{h^0}=A\left(\frac{1}{|v_\chi|^2}+\frac{1}{|v_\rho|^2}+\frac{1}{|v_\eta|^2}\right),
\label{a4}
\end{eqnarray}
where
\begin{eqnarray}
&&N_a=\left(\frac{1}{|v_\chi|^2}+\frac{1}{|v_\eta|^2}\right)^{-1/2},
\quad
N_b=\left(
\frac{1}{|v_\chi|^2}+\frac{|v_\eta|^2}{|v_\rho|^2(|v_\eta|^2+|v_\chi|^2)}\right)^{-1/2},
\nonumber \\&&
N_c=\left(\frac{1}{|v_\chi|^2}+\frac{1}{|v_\rho|^2}+\frac{1}{|v_\eta|^2}\right)^{-1/2}.
\label{a5}
\end{eqnarray}

%\end{itemize}
For the CP-even scalars we are unable to find an analytic solution. But, since the mass matrix is real and symmetric,
we know that it can be diagonalized by an orthogonal matrix. Therefore:  $X^0_\psi=\sum_i O^H_{\psi a}H^0_i$, where
$\psi=\chi,\eta,\rho$, $i=1,2,3$, $H^0_i$ are the mass eigenstates and $O^H$ is an orthogonal matrix.

Notice that since $v_\eta$ and $v_\rho$ are already known in the context of Ref.~\cite{Dias:2006ns} and a lower limit on $\vert v_\chi\vert$ was obtained in Ref.~\cite{Machado:2013jca}, the projection of the scalar symmetry eigenstates over the mass eigenstates is now completely determined. We have used $v_\eta=240$ GeV, $v_\rho=54$ GeV, and $\vert v_\chi\vert=2000$ GeV. 

\section{Lepton-scalar charged interactions}
\label{sec:leptonscalars}

From the Eq.~(\ref{l2}), we obtain the interaction terms of the Lagrangian for the charged leptons and charged scalars:

\begin{eqnarray}
%&&\bar{\nu}_Ll_RY^+_1: -i\frac{\sqrt{2}\vert v_\eta\vert}{\vert v_\rho\vert\sqrt{\vert %v_\eta\vert^2+\vert v_\rho\vert^2}}V^l_L\hat{M}^lP_R,
%\nonumber \\&&
-\mathcal{L}_{E_Ll_RY}=
\bar{E}_L V_{E_Ll_R} l_R Y^{++},
\quad
%\bar{\nu}_L E_RY^-_2:-i\frac{\sqrt{2}\vert v_\rho\vert}{\vert v_\chi\vert \sqrt{\vert v_\rho\vert^2+\vert %v_\chi\vert^2}}\hat{M}e^Ee^{-i\theta_\chi}P_R,\nonumber \\&&
-\mathcal{L}_{\bar{l}_LE_RY}= \bar{l}_L V_{l_LE_R} E_R\,Y^{--},
\label{csi1}
\end{eqnarray}
where 
\begin{equation}
V_{E_Ll_R}=\frac{\sqrt{2}\vert v_\chi\vert}{\vert v_\rho\vert\sqrt{\vert v_\rho\vert^2+\vert v_\chi\vert^2}} V^{l\dagger}_L\hat{M}^l e^{-i\theta_\chi},\;\;V_{l_LE_R}=\frac{\sqrt{2}\vert v_\rho\vert}{\vert v_\chi\vert\sqrt{\vert v_\rho \vert^2+\vert v_\chi\vert^2}} V^l_L\hat{M}^E e^{i\theta_\chi},
\label{l12}
\end{equation}
where $\hat{M}^l$ and $\hat{M}^E$ are, respectively, the diagonal mass matrices of the known leptons $l=e,\mu,\tau$ and the heavy ones $E_e,E_\mu,E_\tau$.
The numerical values of the matrices $V^l_L$ and $V^l_R$ are given in Eqs.~(\ref{vll}) and (\ref{vlr}), respectively. We recall that we have considered a basis in which the heavy leptons mass matrix is diagonal, i.e., that their masses are $m_{E_l}=\vert G^E_{ll}\vert \vert v_\chi\vert/\sqrt{2}$. Otherwise the matrices $V^E_{L,R}$ which diagonalize the general matrix $M^E$ will
appear in the vertexes above. We think that this refinement is not necessary at this time.

\section{Quark-scalar interactions}
\label{sec:quarkscalars}

From Eqs.~(\ref{jjJ}) and (\ref{J}) we obtain the Yukawa interactions with the charged scalars that contribute to the EDM. 

Interactions among $D_L$-type and $J_R$ quarks:
\begin{equation}
-\mathcal{L}_{YD_LJ_R}=\bar{D}_L
K_{D_LJ_R}\,J_R  Y^{--},
\label{ec1}
\end{equation}
where $J_R=(0\,0\,J)_R$ and with
\begin{equation}
K_{D_LJ_R}=\frac{\sqrt{2}e^{-i\theta_\chi}}{\vert v_\chi\vert\sqrt{1+\frac{|v_\chi|^2}{|v_\rho|^2}}}\,V^{D}_L
\left(\begin{array}{ccc}
0 & 0 & 0 \\ 0 & 0 & 0 \\ 0 & 0 & m_J
\end{array}\right).
\label{ec2}
\end{equation}

Interactions among $U_L$-type and $j_R$-type quarks:
\begin{equation}
-\mathcal{L}_{YU_Lj_R}=\bar{U}_L 
K_{U_Lj_R}\,j_RY^{++},
\label{ec5}
\end{equation}
with
\begin{equation}
K_{U_Lj_R}= \frac{\sqrt{2}e^{i\theta_\chi}}{\vert v_\chi\vert\sqrt{1+\frac{|v_\chi|^2}{|v_\rho|^2}}}\,V^{U}_L 
\left( \begin{array}{ccc}
m_{j_1} & 0 & 0 \\ 0 & m_{j_2} & 0 \\ 0 & 0 & 0
\end{array}\right).
\label{ec6}
\end{equation}

Interactions among $J_L$ and $D_R$-type quarks:
\begin{equation}
-\mathcal{L}_{YJ_LD_R}=\bar{J}_L
K_{J_LD_R}\,D_RY^{++},
\label{ec7}
\end{equation}
with
\begin{equation}
K_{J_LD_R}=\frac{|v_\chi|e^{-i\theta_\chi}}{\sqrt{|v_\rho|^2+|v_\chi|^2}}\,\left( \begin{array}{ccc}
0 & 0 & 0 \\ 0 & 0 & 0 \\ \tilde{F}_{31} & \tilde{F}_{32} & \tilde{F}_{33}
\end{array}\right)V_R^{D\dagger}.
\label{ec8}
\end{equation}

Interactions among $j_L$-type and $U_R$-type quarks:
\begin{equation}
-\mathcal{L}_{Yj_LU_R}=\bar{j}_L 
K_{j_LU_R}\,U_R Y^{--},
\label{ec11}
\end{equation}
with
\begin{equation}
K_{j_LU_R}=\frac{|v_\chi|e^{i\theta_\chi}}{\sqrt{|v_\rho|^2+|v_\chi|^2}}\,\left( \begin{array}{ccc}
G_{11} & G_{12} & G_{13} \\ G_{21} & G_{22} & G_{23} \\ 0 & 0 & 0
\end{array}\right)V_R^{U\dagger}.
\label{ec12}
\end{equation}

For the numerical values for the matrices $V^{U,D}_{L,R}$ see Eq.~(\ref{vudl331}) and (\ref{vudr331}) and for those of the parameters in Eqs.~(\ref{ec1}) - (\ref{ec12}) see below Eq.~(\ref{vudr331}).
Notice that both matrices left- and right-handed survive in different interactions in the scalar sector.

\section{Scalar-photon interactions}
\label{sec:verticesgaugescalar}

Now, from the covariant derivatives of the scalar's Lagrangian
\begin{equation}
\mathcal{L}_S=\sum_{i=\eta,\rho,\chi} (D^i\phi_i)^\dagger(D^i\phi_i)
\label{dc}
\end{equation}
where $D^i$ are the covariant derivatives, %in Eq.~(\ref{derivadas_covariantes}),
 we can find the vertexes for the interactions between scalars and photons. The $A_\mu Y^+_{1,2}Y^-_{1,2}$ vertexes are both equal to $ie(k^- -k^+)_\mu$, and the vertex $A_\mu Y^{++}Y^{--}$ is $2ie(k^- -k^+)_\mu $. The terms $k^{+}$ and $k^-$ indicate, respectively, the momenta of the positive and negative charge scalars. The momenta are all going into the vertex and the modulus of the electric charge of the electron is given by
\begin{equation}
e=g\frac{t}{\sqrt{1+4t^2}}=g\sin\theta_W
\end{equation}
with $t=s_W/\sqrt{1-4s^2_W}$.

\newpage

\newpage

\newpage

\newpage

%\begin{figure}[tbp]
%\centering % \begin{center}/\end{center} takes some additional vertical space
%\includegraphics[width=.75\textwidth,origin=c,angle=0]{./EDM_neutron_mYPP}
%\hfill
%\caption{Allowed values for $\theta_\chi$ and $m_{Y^{++}}$ from the neutron EDM.
%The shaded regions show the values for $\theta_\chi$ and $m_{Y^{++}}$
%where $|d_{n331}|<2.9\times 10^{-26}$ e cm, see Eqs.
%(\ref{eq:mde_331}) - (\ref{edmn2})). In this plot we considered
%$m_{Y_2^+}=300$ GeV, $|v_\chi|=2$ TeV and 1 TeV for the masses of the
%exotic quarks. From this graph it is seen that all values of
%$m_{Y^{++}}$ are allowed for the values of $\theta_\chi$ where
%$\sin(2\theta_\chi)=0$ (for some of these values the plot resolution
%does not show properly the allowed region). For the phase values where
%the sine is not null, it can be seen that larger values for
%$m_{Y^{++}}$ are favored, an indication that the EDM decreases as the
%mass of the scalar increases. }
%\label{fig:edmneutron1}
%\end{figure}


\begin{thebibliography}{99}



%\cite{Beringer:1900zz}
%\bibitem{Beringer:1900zz}
%  J.~Beringer {\it et al.}  [Particle Data Group Collaboration],
%  %``Review of Particle Physics (RPP),''
%  Phys.\ Rev.\ D {\bf 86}, 010001 (2012).

\bibitem{Shabalin:1978}
E.~P.~Shabalin.
%"Electric dipole moment of the quark in a gauge theory with left-handed currents,"
Sov.\ J.\ Nucl.\ Phys.\ {\bf 28}, 75 (1978).

\bibitem{Shabalin:1983}
E.~P.~Shabalin.
%"Electric dipole moment of the neutron in gauge theory,"
Sov.\ Phys.\ Usp\ {\bf 26}, 297 (1983).

\bibitem{Shabalin:1980}
E.~P.~Shabalin.
%"Baryon Electric Dipole Moments in CP-noninvariant Kobayashi-Maskawa Theory,"
Sov.\ J.\ Nucl.\ Phys.\ {\bf 32}, 228 (1980).

\bibitem{Eeg:1984}
J.~O.~Eeg, I.~Picek.
%"Two-loop diagrams for the electric dipole moment of the neutron,"
Nucl.\ Phys.\  {\bf B244}, 77 (1984).

\bibitem{Czarnecki:1997}
A.~Czarnecki, B.~Krause.
%"Neutron Electric Dipole Moment in the Standard Model: Complete Three-Loop Calculation of the Valence Quark Contributions,"
Phys.\ Rev.\ Lett.\ \textbf{78}, 4339 (1997).


\bibitem{Baker:2006}
C.~Baker, D.~D.~Doyle, P.~Geltenbort, {\it et al},
%"Improved Experimental Limit on the Electric Dipole Moment of the Neutron,"
Phys.\ Rev.\ Lett.\ {\bf 97}, 131801 (2006).

\bibitem{Cummins:1999}
E.~D.~Cummins,
%"Electric Dipole Moments of Leptons"
Adv. Atom. Mol. Opt. Phys., {\bf 40}, 1 (1999).

%\cite{Baron:2013eja}
\bibitem{Baron:2013eja}
  J.~Baron {\it et al.}  (ACME Collaboration),
  %``Order of Magnitude Smaller Limit on the Electric Dipole Moment of the Electron,''
  Science {\bf 343}, 269 (2014).

\bibitem{Riotto:1999}
A.~Riotto and M.~Trodden,
%"Recent Progress in Baryogenesis,"
Ann.\ Rev.\  Nucl.\ Part.\ Sci.\ {\bf 49}, 35 (1999).

\bibitem{Morrisey:2012}
D.~E.~Morrissey and M.~J.~Ramsey-Musolf,
%"Electroweak baryogenesis,"
New\ J.\ Phys.\ {\bf14}, 125003 (2012).

\bibitem{Dine:2004}
M.~Dine and A.~Kusenko,
%"Origin of the matter-antimatter asymmetry,"
Rev.\ Mod.\ Phys.\ {\bf 76}, 1 (2004).

%\cite{Pleitez:1992xh}
\bibitem{Pleitez:1992xh}
  V.~Pleitez and M.~D.~Tonasse,
  %``Heavy charged leptons in an SU(3)-L x U(1)-N model,''
  Phys.\ Rev.\ D {\bf 48}, 2353 (1993).

%\cite{Promberger:2007py}
\bibitem{Promberger:2007py}
  C.~Promberger, S.~Schatt and F.~Schwab,
  %``Flavor Changing Neutral Current Effects and CP Violation in the Minimal 3-3-1 Model,''
  Phys.\ Rev.\ D {\bf 75}, 115007 (2007).


%\cite{Montero:1998yw}
\bibitem{Montero:1998yw}
  J.~C.~Montero, V.~Pleitez and O.~Ravinez,
  %``Soft superweak CP violation in a 331 model,''
  Phys.\ Rev.\ D {\bf 60}, 076003 (1999).

%\cite{Montero:2005yb}
\bibitem{Montero:2005yb}
J.~C.~Montero, C.~C.~Nishi, V.~Pleitez, O.~Ravinez and M.~C.~Rodriguez,
%``Soft CP violation in $K$ meson systems,''
Phys.\ Rev.\ D {\bf 73}, 016003 (2006).

%\cite{Agashe:2014kda}
\bibitem{Agashe:2014kda} 
  K.~A.~Olive {\it et al.}  [Particle Data Group Collaboration],
  %``Review of Particle Physics,''
  Chin.\ Phys.\ C {\bf 38}, 090001 (2014).

\bibitem{Machado:2013jca}
A.~C.~B.~Machado, J.~C.~Montero and V.~Pleitez,
%``FCNC in the minimal 3-3-1 model revisited,''
Phys.\ Rev.\ D {\bf 88}, 113002 (2013).


\bibitem{Dias:2006ns} A.~G.~Dias, J.~C.~Montero and V.~Pleitez,
%  \textsl{Closing the $SU(3)_L \otimes  U(1)_X$ symmetry at electroweak scale},
Phys.\ Rev.\ D {\bf 73}, 113004 (2006).

%\bibitem{leptons331} A. C. B. Machado \textit{et al.} work in progress.

%\cite{Liu:1993gy} 
\bibitem{Liu:1993gy} 
  J.~T.~Liu and D.~Ng,
  %``Lepton flavor changing processes and CP violation in the 331 model,''
  Phys.\ Rev.\ D {\bf 50}, 548 (1994).

%\cite{Davey:2014tka}
\bibitem{Davey:2014tka} 
  W.~Davey,
  %``BSM Higgs boson searches at LHC and the Tevatron,''
  arXiv:1409.6016.

%\cite{Pospelov:2005pr}
\bibitem{Pospelov:2005pr} 
  M.~Pospelov and A.~Ritz,
  %``Electric dipole moments as probes of new physics,''
  Annals Phys. (Berlin)  {\bf 318}, 119 (2005).
%\cite{Ritz:2009zz}
\bibitem{Ritz:2009zz} 
  A.~Ritz,
  %``Probing new CP-odd physics with electric dipole moments,''
  Nucl.\ Instrum.\ Meth.\ A {\bf 611}, 117 (2009).


\end{thebibliography}
\end{document}